\documentclass{SolarPhysics}    
\usepackage[optionalrh]{spr-sola-addons}
\usepackage{epsfig}
\usepackage{url}
\usepackage{fixltx2e}
\usepackage[usenames]{color}

\newcommand{\gapprox}{\lower.4ex\hbox{$\;\buildrel >\over{\scriptstyle\sim}\;$}}
\newcommand{\lapprox}{\lower.4ex\hbox{$\;\buildrel <\over{\scriptstyle\sim}\;$}}
\newcommand{\arcsec}{\hbox{$^{\prime\prime}$}}

\def\etal{{\it et al.,~}}
\def\eg{{\it e.g.,~}}
\def\ie{{\it i.e.,~}}
\def\etc{{\it etc.}}

\def\aap  {{\sl Astron. Astrophys.}\ }   
\def\apj  {{\sl Astrophys. J.}\ }        
\def\apjl {{\sl Astrophys. J. Lett.}\ }  
\def\aj   {{\sl Astronom. J.}\ } 	 
\def\mnras{{\sl Mon.~Not.~Roy.~Astron.~Soc.}\ } 
\def\sp   {{\sl Solar Phys.}\ }          
\def\ssr  {{\sl Space Sci. Rev.}\ }      

\begin{document}
\begin{article}
\begin{opening}
\title{A Nonlinear Force-Free Magnetic Field Approximation 
	Suitable for Fast Forward-Fitting to Coronal Loops. I. Theory}

\author{Markus J. Aschwanden}
\runningauthor{M.J. Aschwanden}
\runningtitle{Nonlinear Force-Free Magnetic Field}

\institute{Solar and Astrophysics Laboratory,
	Lockheed Martin Advanced Technology Center, 
        Dept. ADBS, Bldg.252, 3251 Hanover St., Palo Alto, CA 94304, USA; 
        (e-mail: \url{aschwanden@lmsal.com})}

\date{Received 30 Nov 2011; Revised 1 July 2012; Accepted ...}

\begin{abstract}
We derive an analytical approximation of nonlinear force-free 
magnetic field solutions (NLFFF) that can efficiently be used 
for fast forward-fitting to solar magnetic data, constrained either
by observed line-of-sight magnetograms and stereoscopically
triangulated coronal loops, or by 3D vector-magnetograph data.
The derived NLFFF solutions provide the magnetic field components 
$B_x({\bf x})$, $B_y({\bf x})$, $B_z({\bf x})$, the force-free 
parameter $\alpha({\bf x})$, the electric current density 
${\bf j}({\bf x})$, and are accurate to second-order (of the
nonlinear force-free $\alpha$-parameter). 
The explicit expressions of a force-free field can easily be applied 
to modeling or forward-fitting of many coronal phenomena. 
\end{abstract}


\end{opening}

\section{		Introduction 			}

The coronal magnetic field can be constrained in a number of ways,
such as by extrapolation of photospheric magnetograms or 
vector-magnetograph data, by radio observations of gyroresonance
layers above sunspots, of by coronal seismology of oscillating loops.
Before the advent of the STEREO mission, attempts were made to
model observed coronal loops with stretched potential field solutions
(Gary and Alexander, 1999), to fit a linear force-free model with 
solar-rotation stereoscopy (Wiegelmann and Neukirch, 2002; Feng \etal 2007a), 
by tomographic reconstruction with magnetohydrostatic constraints 
(Wiegelmann and Inhester, 2003; Ruan \etal 2008), by magnetic modeling
applied to spectropolarimetric loop detections (Wiegelmann \etal 2005), 
or by magnetic field supported stereoscopic loop triangulation 
(Wiegelmann and Inhester, 2006; Conlon and Gallagher, 2010).
Recently, stereoscopic triangulation of coronal loops with the STEREO
mission became available, which constrains the 3D geometry of coronal
magnetic field lines (Aschwanden \etal 2008; Aschwanden, 2009). 
The plethora of coronal high-resolution data
allows us now to compare different magnetic models and to test whether
they are self-consistent. A critical assessment of nonlinear force-free
field (NLFFF) codes revealed the disturbing fact that different
NLFFF codes yield incompatible results among themselves, and exhibit 
significant misalignments with stereoscopically triangulated loops 
(DeRosa \etal 2009;
Sandman \etal 2009; Aschwanden and Sandman, 2010; Sandman and Aschwanden,
2010; Aschwanden \etal 2012a,b). The discrepancy was attributed to 
uncertainties in the boundary conditions as well as to the
non-forcefreeness of the photosphere and lower chromosphere. Earlier
tests with the virial theorem already indicated that the magnetic fields
in the lower chromosphere at altitudes of $h \lapprox 400$ km are not
force-free (Metcalf \etal 1996). Constraints by coronal tracers thus
have become an important criterion to bootstrap a self-consistent magnetic 
field solution. The misalignment between theoretical extrapolation models
and stereoscopically triangulated loops could be minimized by using
potential field models with forward-fitted unipolar magnetic charges
(Aschwanden and Sandman, 2010) or dipoles (Sandman and Aschwanden, 2011).

In this Paper we go a step further by deriving a simple analytical
approximation of nonlinear force-free field solutions that is suitable
for fast forward-fitting to stereoscopically triangulated loops
or to some other coronal observations.
While accurate solutions of force-free magnetic fields have been known
for special mathematical functions (Low and Lou, 1990) that have been
used to reconstruct the local twist of coronal loops 
(Malanushenko \etal 2009, 2011), they are not suitable for forward-fitting
to entire active regions. In contrast, our theoretical framework entails
the representation of a potential or non-potential field by a superposition
of a finite number of elementary field components that are associated with
buried unipolar magnetic charges at arbitrary locations, each one being
divergence-free and force-free to a good approximation, as we test 
numerically. While this Paper contains the analytical framework of the
magnetic field model, the numerical forward-fitting code with applications
to observations will be presented in a Paper II (Aschwanden and
Malanushenko, 2012), and applications to stereoscopically observed 
active regions in Aschwanden \etal (2012a,b).

\section{		Theory 				}

\subsection{	Potential Field Parameterization	}

The simplest representation of a magnetic potential field
that fulfills Maxwell's divergence-free condition ($\nabla \cdot {\bf B}=0$)
is a unipolar magnetic charge $j$ that is buried below the solar surface,
which predicts a magnetic field
${\bf B}_j({\bf x})$ that points away from the buried unipolar charge
and whose field strength falls off with the square of the distance $r_j$,
\begin{equation}
        {\bf B_j}({\bf x})
        = B_j \left({d_j \over r_j}\right)^2 {{\bf r}_j \over r_j} \ ,
\end{equation}
where $B_j$ is the magnetic field strength at the solar surface 
above a buried magnetic charge, $(x_j, y_j, z_j)$ is the
subphotospheric position of the buried charge, $d_j$ is the depth of 
the magnetic charge, 
\begin{equation}
	d_j = 1-\sqrt{x_j^2+y_j^2+z_j^2} \ ,
\end{equation}
and ${\bf r}_j=[x-x_j, y-y_j, z-z_j]$ is the vector between an arbitrary 
location ${\bf x}=(x,y,z)$ in the solar corona (were we desire to calculate 
the magnetic field) and the location $(x_j, y_j, z_j)$ of the buried charge. 
We choose a Cartesian coordinate system
$(x,y,z)$ with the origin in the Sun center and are using units of solar
radii, with the direction of $z$ chosen along the line-of-sight from Earth
to Sun center. For a location near disk center ($x \ll 1, y \ll 1$), the
magnetic charge depth is $d_j \approx (1-z_j)$.
Thus, the distance $r_j$ from the magnetic charge is 
\begin{equation}
	r_j = \sqrt{(x-x_j)^2+(y-y_j)^2+(z-z_j)^2} \ .
\end{equation}
The absolute value of the magnetic field $B_j(r_j)$ is simply a function of
the radial distance $r_j$ (with $B_j$ and $d_j$ being constants for a
given magnetic charge),
\begin{equation}
	B(r_j) = B_j \left({d_j \over r_j}\right)^2 \ .
\end{equation}

In order to obtain the Cartesian coordinates
$(B_x, B_y, B_z)$ of the magnetic field vector ${\bf B}_j({\bf x})$, 
we can rewrite Equation (1) as,
\begin{equation}
	\begin{array}{ll}
		B_x(x,y,z) &= B_j \left({d_j / r_j}\right)^2 
			     (x-x_j) / r_j \\
		B_y(x,y,z) &= B_j \left({d_j / r_j}\right)^2 
			     (y-y_j) / r_j \\
		B_z(x,y,z) &= B_j \left({d_j / r_j}\right)^2 
			     (z-z_j) / r_j \\
	\end{array} \ .
\end{equation}

\begin{figure}
\centerline{\includegraphics[width=0.8\textwidth]{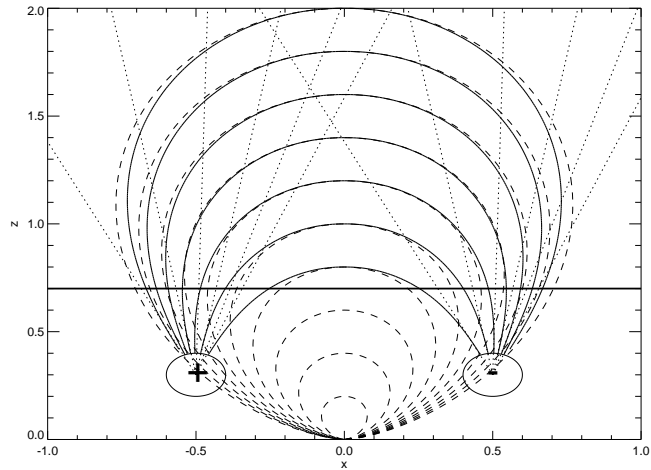}}
\caption{The magnetic field of a symmetric dipole (dashed lines)
is shown, together with the field resulting from the superposition
of two unipolar magnetic charges (solid lines). The two field models
become identical once the two unipolar charges are moved towards the
location of the dipole moment at position $(x,y)=(0,0)$. The radial
field of each unipolar (positive and negative) charge is also shown
for comparison (dotted lines).}
\end{figure}

We progress now from a single magnetic charge to an arbitrary number
$N_{\rm m}$ of magnetic charges and represent the general magnetic field with a
superposition of $N_{\rm m}$ buried magnetic charges, so that the potential field 
can be represented by the superposition of $N_{\rm m}$ fields ${\bf B}_j$ from 
each magnetic charge $j=1,...,N_{\rm m}$,
\begin{equation}
        {\bf B}({\bf x}) = \sum_{j=1}^{N_{\rm m}} {\bf B}_j({\bf x})
        = \sum_{j=1}^{N_{\rm m}}  B_j
        \left({d_j \over r_j}\right)^2 {{\bf r_j} \over r_j} \ .
\end{equation}

As an example we show the representation of a dipole with two magnetic
unipolar charges ($N_{\rm m}=2$) of opposite polarity ($B_2=-B_1$) in Figure 1.
Each of the unipolar charges has a radial magnetic field (dotted lines), 
but the superposition of the two vectors of both unipolar charges in
every point of space, ${\bf B}({\bf x})={\bf B_1}({\bf x}) 
+{\bf B_2}({\bf x})$, reproduces the familiar dipole field. For the case 
shown in Figure 1 we used the parameterization of two subphotospheric unipolar 
magnetic charges at positions $x_1=-0.5$ and $x_2=+0.5$, which produces 
dipole-like field lines (solid curves), while they converge to the 
classical solution of a dipole field in the limit of 
$x_1 \mapsto 0$ and $x_2 \mapsto 0$,
as it can be shown analytically (Jackson, 1972; p.184). 

\subsection{	Force-Free Field Solution of a Uniformly Twisted Fluxtube }

A common geometrical concept is to characterize coronal loops with cylindrical
fluxtubes. For thin fluxtubes, the curvature of coronal loops and the
related forces can be neglected, so that a cylindrical geometry can be applied.
Because the footpoints of coronal loops are anchored in the photosphere, where
a random velocity field creates vortical motion on the coronal fluxtubes,
they are generally twisted. We consider now such twisted fluxtubes
in a cylindrical geometry and derive a relation between the helical twist 
and the force-free parameter $\alpha$. The analytical solution of a uniformly
twisted flux tube is described in several textbooks (\eg Gold and Hoyle, 1960;
Priest, 1982; Sturrock, 1994; Boyd and Sanderson, 2003; 
Aschwanden, 2004), but we summarize the derivation here 
to provide physical insights for the generalized derivation of nonlinear 
force-free magnetic field solutions derived in Section 2.3 
in a self-consistent notation. 

\begin{figure}
\centerline{\includegraphics[width=0.7\textwidth]{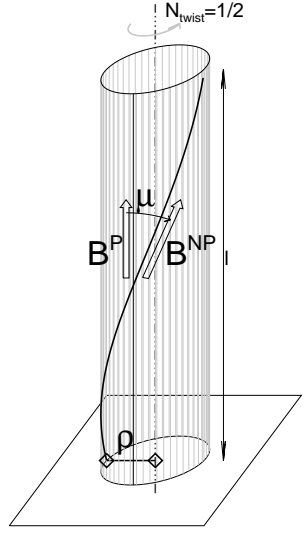}}
\caption{The basic 3D geometry of a cylindrical flux tube with uniform
twist is defined by the length $l$ of the cylinder axis, the number
of twisting turns along this length, $N_{\rm twist}$, or by the misalignment
angle $\mu$ at the flux tube radius $\rho$ between the potential field line
${\bf B}^{\rm P}$ (aligned with the cylindrical axis) and the non-potential
field line ${\bf B}^{\rm NP}$ (aligned with the twisted loop). The
non-potential field line ${\bf B}^{\rm NP}$ can be decomposed into a 
longitudinal
field component $B_s$ and an azimuthal field component $B_{\varphi}$.}
\end{figure}

We consider a straight cylinder where a uniform twist is applied,
so that an initially straight field line ${\bf B}=(0, 0, B_s)$,
aligned with a field line coordinate $s$, is rotated by a
number $N_{\rm twist}$ of full turns over the cylinder length $l$, 
yielding an azimuthal field component $B_{\varphi}$ at radius $\rho$,
\begin{equation}
        {B_{\varphi} \over B_s} 
	= { \rho \partial \varphi \over \partial s} 
	= { 2 \pi \rho N_{\rm twist} \over l} = b \rho \ ,
\end{equation}
with the constant $b$ defined in terms of the number of full twisting turns 
$N_{\rm twist}$ over a (loop) length $l$.  

The cylindrical geometry of a twisted flux tube is
visualized in Figure 2. The longitudinal component of the untwisted magnetic 
field corresponds to a potential field vector ${\bf B}^{\rm P}$, 
while the twisted 
non-potential field line ${\bf B}^{\rm NP}$ has a helical geometry with an
angle $\mu$ at a radius $\rho$, which can be described by the longitudinal
component $B_s$ and the azimuthal component $B_{\varphi}$. 
The fluxtube can be considered as a sequence of cylinders with radii $\rho$, 
each one twisted by the same twist angle $\partial \varphi/\partial s=2 
\pi N_{\rm twist}/l$. For uniform twisting, the magnetic components 
$B_{\varphi}$ and $B_s$ depend only on the 
radius $\rho$, but not on the length coordinate $s$ or azimuth angle $\varphi$. 
Thus, the functional dependence in cylindrical coordinates 
$(\rho, \varphi, s)$ is
\begin{equation}
        {\bf B} = \left[ B_{\rho}, B_{\varphi}, B_s \right] =
                  \left[ 0  , B_{\varphi}(\rho), B_s(\rho) \right]  \ .
\end{equation}
Consequently, the general expression of $\nabla \times {\bf B}$ in cylindrical 
coordinates,
\begin{equation}
        {\nabla} \times {\bf B} =
        \left[ {1 \over \rho} {\partial B_s \over \partial \varphi} - 
	{\partial B_{\varphi} \over \partial s} ,
        {\partial B_\rho \over \partial s} - {\partial B_s \over \partial \rho} ,
        {1 \over \rho} \left( {\partial \over \partial \rho}(\rho B_{\varphi}) - 
	{\partial B_\rho \over \partial \varphi}\right) \right] \ ,
\end{equation}
is simplified with $B_{\rho}=0$ and the sole dependencies of
$B_{\varphi}(\rho)$ and $B_s(\rho)$ on the radius $\rho$ (Equation (7)), 
yielding a force-free current density ${\bf j}$ of,
\begin{equation}
        {\bf j} = \left[ j_\rho, j_{\varphi}, j_s \right]
                = {c \over 4\pi} ({\nabla} \times {\bf B})
                = {c \over 4\pi} \left[ 0, - {\partial B_s \over \partial \rho},
                {1 \over \rho} \left( {\partial \over \partial \rho}
		(\rho B_{\varphi})\right) \right] \ .
\end{equation}
Requiring that the Lorentz force is zero for a force-free solution, 
${\bf F}={\bf j} \times {\bf B} = 0$, we obtain a single non-zero component 
in the radial $\rho$-direction, since $j_\rho=0$ and $B_{\rho}=0$ for the 
two other components,
\begin{equation}
        {\bf F} = {\bf j} \times {\bf B} 
	= \left[ B_s j_{\varphi} - B_{\varphi} j_s, 0, 0 \right] \ ,
\end{equation}
yielding a single differential equation for $B_s$ and $B_{\varphi}$,
\begin{equation}
        B_s {dB_s \over d\rho} + B_{\varphi} {1 \over \rho} {d \over d\rho} 
	(\rho B_{\varphi}) = 0 \ .
\end{equation}
By substituting $B_{\varphi}=b \rho B_s$ from Equation (7) into Equation (12), 
this simplifies to, 
\begin{equation}
        {d \over d\rho} \left[ ( 1 + b^2 \rho^2) B_s \right] = 0 \ .
\end{equation}
A solution is found by making the expression inside the derivative to 
a constant ($B_0$), which yields $B_{\varphi}$ and $B_s$,
\begin{equation}
        {\bf B} = \left[ B_\rho, B_{\varphi}, B_s \right] =
        \left[ 0, {B_0 \ b \rho \over 1 + b^2 \rho^2}, {B_0 \over 1 + b^2 \rho^2} 
	\right] \ .
\end{equation}
[This equation corrects also a misprint in Equation (5.5.8) 
of Aschwanden (2004; p.216),
where a superflous zero component has to be eliminated]. 
With the definition of the force-free $\alpha$-parameter,
\begin{equation}
       (\nabla \times {\bf B}) = {4\pi \over c} {\bf j} 
	= \alpha (\rho) {\bf B} \ ,
\end{equation}
we can now verify that the $\alpha$-parameter 
for a uniformly twisted fluxtube depends only on the radius $\rho$,
\begin{equation}
        \alpha (\rho) = {2 b \over (1 + b^2 \rho^2)} \ , 
\end{equation}
with the constant $b$ defined in terms of the number of full twisting turns 
$N_{\rm twist}$ over a (loop) length $l$ (see Equation (7)),  
\begin{equation}
	b = {2 \pi N_{\rm twist} \over l} \ .
\end{equation}

\begin{figure}
\centerline{\includegraphics[width=0.8\textwidth]{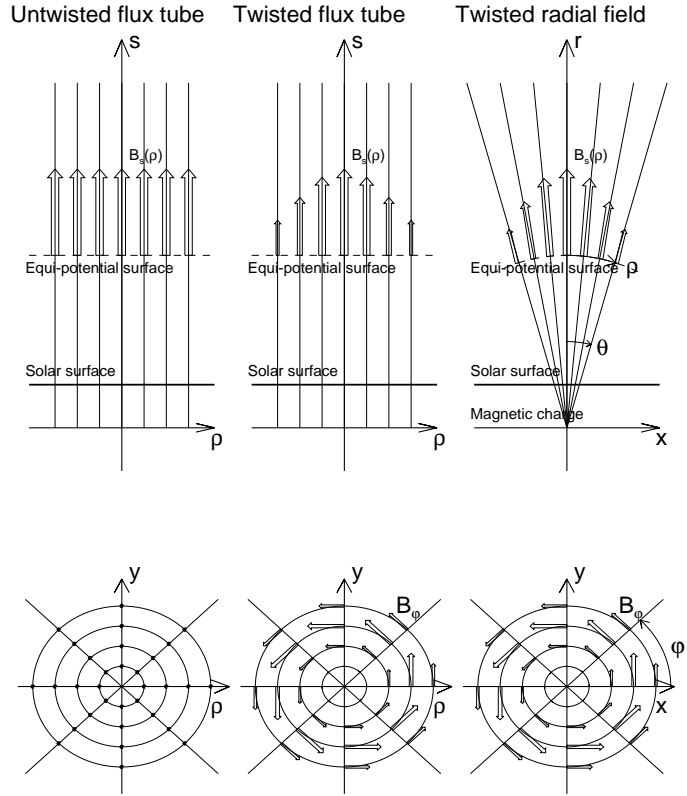}}
\caption{The field line geometry is shown for an untwisted
cylindrical flux tube (left), a twisted cylindrical flux tube (middle),
and for a twisted radial field (right), from the side view in the
$xz$-plane (top) and from the top view in the $xy$-plane (bottom).
The top panels show the longitudinal magnetic field component $B_s(\rho)$ 
and the bottom panels show the azimuthal magnetic field component
$B_{\varphi}(\rho, \varphi)$.} 
\end{figure}

The geometry of a twisted flux tube is visualized in Figure 3 (top middle),
where the parallel field lines are aligned with the coordinate axis $s$ 
in the vertical direction, the cross-sectional radius $\rho$ is defined in 
the direction perpendicular to $s$, and the twist angle $\varphi$ is
indicated in the horizontal projection (Figure 3, bottom middle). 
According to Equations (8) and (14), the variations of the longitudinal 
$B_s(\rho)$
and of the azimuthal component $B_{\varphi}(\rho)$ with radius $\rho$ are,
\begin{equation}
	B_s(\rho) = {B_0 \over 1 + b^2 \rho^2} \ ,
\end{equation}
\begin{equation}
	B_{\varphi}(\rho) = {B_0 b \rho \over 1 + b^2 \rho^2} \ .
\end{equation}
These radial dependencies are shown in Figure 4 for different
numbers of twist ($N_{\rm twist}=0.5, 1.0, 1.5$).
In the limit of vanishing twist ($N_{\rm twist}=0 \mapsto b =0$),
we have an untwisted flux tube (Figure 3 left) with a constant
longitudinal field $B_s(\rho)=B_0$ and a vanishing azimuthal component
$B_{\varphi}(\rho)=0$. 
The dependence of the azimuthal field component $B_{\varphi}(\rho)$
and the longitudinal field component $B_s(\rho)$
as a function of the radius ${\rho}$ from the twist axis 
(Figure 4) shows that the longitudinal
component falls off monotonically with radius $\rho$, while the azimuthal 
component increases first for small distances $\rho \ll l$, but falls off 
at larger distances. Thus, the twisting causes a smaller cross-section of 
a fluxtube compared with the potential field situation, as widely known
(\eg Klimchuk \etal 2000).

\begin{figure}
\centerline{\includegraphics[width=0.8\textwidth]{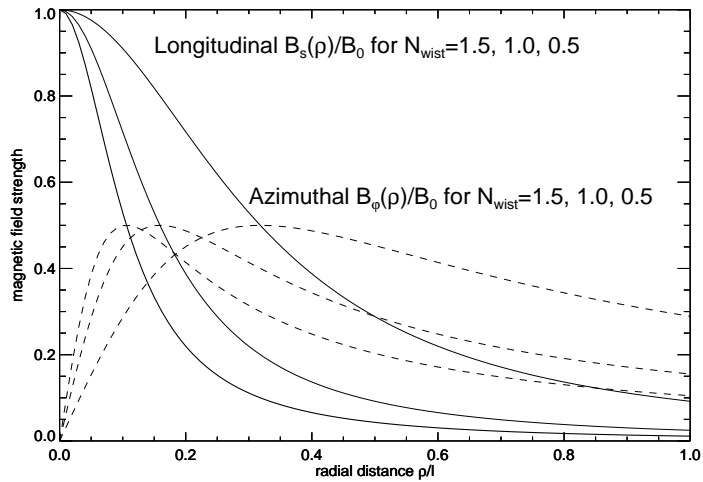}}
\caption{The dependence of the longitudinal (solid lines) and
azimuthal magnetic field component (dashed lines) as a function
of the distance $r/l$ from the twist axis field is shown for
three different amounts of twist ($N_{\rm twist}=1.5, 1.0, 0.5$ full
turns per loop length $l$).}
\end{figure}

\subsection{	Nonlinear Force-Free Field Parameterization	}

We are now synthesizing the concept of point-like buried magnetic
charges that we used to parameterize a potential field (Section 2.1)
with the uniformly twisted flux tube concept that represents an exact
solution of a nonlinear force-free field (Section 2.2). The geometric
difference between the two concepts is the spherical symmetry of
a point charge versus the parallel field configuration of an untwisted
flux tube. However, we can synthesize the two geometries by considering
the parallel field as a far-field approximation of a radial field.
In an Euclidean parallel field, the equi-potential surface is a plane
perpendicular to the parallel field vector, while a radial field has spherical
equi-potential surface. We can make the transformation of a parallel
field in cylindrical coordinates $(s, \rho, \varphi)$ into a radial field 
with spherical coordinates $(r, \theta, \varphi)$ by mapping (see Figure 3), 
\begin{equation}
	\begin{array}{ll}
	s    & \mapsto r \\
	\rho & \mapsto r \sin(\theta) \\
	\end{array} \ .
\end{equation}
This transformation from cylindrical to spherical coordinates preserves
the orthogonality of the longitudinal field component $(B_s \mapsto B_r)$
to the equi-potential surface ($s=$const $\mapsto r=$const) and conserves
the magnetic flux $\Phi(r)$ along a bundle of field lines with area 
$A(r)={\rho^2(r)}$,
\begin{equation}
	\Phi(r) = B(r) A(r) = B(r) \rho^2(r) = B(r) r^2 \sin^2{\theta} 
	= {\rm const} \ ,
\end{equation}
if the longitudinal component $B(r) \propto r^{-2}$ (Equation (1)) decreases
quadratically with distance from the magnetic charge. Thus, applying
the transformation into spherical coordinates (Equation (20)) and the magnetic
flux conservation (Equation (21)) to the straight flux tube solution 
(Equations (18) and (19)), 
we can already guess the approximate nonlinear force-free solution 
in spherical coordinates,
\begin{equation}
	B_r(r, \theta) \propto r^{-2}
	{1 \over (1 + b^2 r^2 \sin^2{\theta})} \ , 
\end{equation}
\begin{equation}
	B_\varphi(r, \theta) \propto r^{-2}
	{b r \sin{\theta} \over (1 + b^2 r^2 \sin^2{\theta})} \ .
\end{equation}

More rigorously, 
we can derive a nonlinear force-free field solution by writing the
divergence-free condition $(\nabla \cdot {\bf B}) = 0$ and the force-free 
condition $(\nabla \times {\bf B}) = (4\pi/c) {\bf j} = \alpha (\rho) {\bf B}$
(Equation (15)) of a magnetic field vector ($B_r, B_\theta, B_\varphi)$ 
in spherical coordinates $(r, \theta, \varphi)$ (with the origin at the
location of the magnetic charge and the spherical symmetry axis aligned 
with the vertical direction to the local solar surface),
\begin{equation}
       (\nabla \cdot {\bf B}) = {1 \over r^2} {\partial \over \partial r}
	(r^2 B_r)  
	+ {1 \over r \sin{\theta}} {\partial \over \partial \theta}
	(B_\theta \sin{\theta}) 
	+ {1 \over r \sin{\theta}}
	{\partial B_\varphi \over \partial \varphi} = 0 \ ,
\end{equation}
\begin{equation}
	\left[ \nabla \times {\bf B} \right]_r =
 	{1 \over r \sin{\theta}}
 	\left[{\partial \over \partial \theta}
 	(B_\varphi \sin{\theta}) -
	{\partial B_\varphi \over \partial \varphi} \right]
	= \alpha B_r \ ,
\end{equation}
\begin{equation}
	\left[ \nabla \times {\bf B} \right]_\theta =
 	{1 \over r}
 	\left[{1 \over \sin{\theta}} {\partial B_r \over \partial \varphi}
	-{\partial \over \partial r} (r B_\varphi) \right]
	= \alpha B_\theta \ ,
\end{equation}
\begin{equation}
	\left[ \nabla \times {\bf B} \right]_\varphi =
 	{1 \over r}
 	\left[{\partial \over \partial r}( r B_\theta ) 
	-{\partial B_r \over \partial \theta} \right]
	= \alpha B_\varphi \ .
\end{equation}
For a simple approximative nonlinear force-free solution we 
require axi-symmetry with no azimuthal dependence 
($\partial / \partial \varphi = 0$) and neglect components
that contribute only to second order ($B_\theta \propto 
[b r \sin{\theta}]^2 \approx 0)$, in analogy to the uniformly twisted 
flux tubes on cylindrical surfaces (Figure 2).
This requirement simplifies Equations (24)--(27) to, 
\begin{equation}
        {1 \over r^2} {\partial \over \partial r} (r^2 B_r) \approx 0 \ ,
\end{equation}
\begin{equation}
 	{1 \over r \sin{\theta}}
	{\partial \over \partial \theta}(B_\varphi \sin \theta) = \alpha B_r \ ,
\end{equation}
\begin{equation}
 	-{1 \over r} {\partial \over \partial r} (r B_\varphi) 
	\approx 0 \ ,
\end{equation}
\begin{equation}
 	- {1 \over r} {\partial B_r \over \partial \theta} 
	\approx \alpha B_\varphi \ .
\end{equation}
Eliminating $\alpha$ from Equations (29) and (31) and using the analog 
{\sl ansatz} as for cylindrical fluxtubes (Equation (7)),
\begin{equation}
	B_\varphi = B_r b r \sin{\theta} \ ,
\end{equation}
we obtain a similar differential equation as in Equation (13),
\begin{equation}
	{\partial \over \partial \theta}
	\left[ B_r ( 1 + b^2 r^2 \sin^2{\theta} ) \right] = 0 \ .
\end{equation}
A solution of this differential equation is obtained by setting the 
expression inside the bracket to the constant $B_0 (d^2 /  r^2)$,
which fulfills the divergence-free condition (Equation (28)), and
we obtain a solution for $B_r$ and $B_\varphi$ (using Equation (32)),
for $\alpha$ (using Equation (29)), 
\begin{equation}
	B_r(r, \theta) = B_0 \left({d^2 \over r^2}\right)
	{1 \over (1 + b^2 r^2 \sin^2{\theta})} \ , 
\end{equation}
\begin{equation}
	B_\varphi(r, \theta) = 
	B_0 \left({d^2 \over r^2}\right)
	{b r \sin{\theta} \over (1 + b^2 r^2 \sin^2{\theta})} \ ,
\end{equation}
\begin{equation}
	B_\theta(r, \theta) \approx 0
	\ ,
\end{equation}
\begin{equation}
	\alpha(r, \theta) \approx {2 b \cos{\theta} \over 
	(1 + b^2 r^2 \sin^2{\theta})}  \ .
\end{equation}
This solution fulfills both the force-free condition (Equations (29)--(31)) 
and the divergence-free condition (Equation (28)) to second-order accuracy
($\propto [b r \sin{\theta}]^2$).  We see that this solution is
identical with the simplified derivation of Equations (22) and (23).
At locations near the twist axis ($\theta \mapsto 0$), the general solution
(Equations (34)--(37)) converges to the cylindrical flux tube geometry solution
(Equations (18) and (19)). Furthermore, in the limit of vanishing twist 
($b \mapsto 0$) 
we retrieve the potential-field solution (Equation (4)), since
the force-free parameter becomes $\alpha \mapsto 0$, the azimuthal 
field component becomes $B_\varphi=0$, and the radial component 
reproduces the potential-field solution $B_r \mapsto B_0 (d^2 / r^2)$.

\begin{figure}
\centerline{\includegraphics[width=0.6\textwidth]{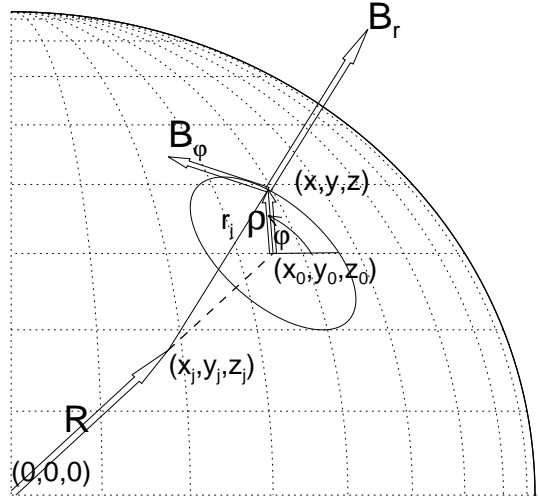}}
\caption{The geometry of a twisted radial field of a magnetic
charge $j$ buried at a subphotospheric position $(x_j,y_j,z_j)$ is shown.
The central twist axis (dashed line) intersects an equi-potential surface 
at position $(x_0, y_0, z_0)$ and the longitudinal field vector $B_r$ 
at position $(x,y,z)$ has a radial distance $\rho$ from the twist axis
and an azimuth angle $\varphi$. The azimuthal magnetic field component 
$B_{\varphi}$ at location $(x,y,z)$ is orthogonal to the radial vector 
${\rho}$ and the longitudinal field component $B_r$, as well as to the
direction of the twist axis ${\bf R}$.}
\end{figure}

\subsection{	Cartesian Coordinate Transformation 		}

In the derivation in the last section we derived the solution in terms
of spherical coordinates $(r, \theta, \varphi)$ in a coordinate system
where the rotational symmetry axis is aligned with the vertical to the
solar surface intersecting a magnetic charge $j$. Since we are going
to model a number of magnetic charges at arbitrary positions on the
solar disk, we have to transform an individual coordinate system
$(r_j, \theta_j, \varphi_j)$ associated with magnetic charge $j$
into a Cartesian coordinate system $(x,y,z)$
that is given by the observers line-of-sight (in $z$-direction) and
the observer's image coordinate system $(x,y)$ in the plane-of-sky.
The variables of the Cartesian coordinate transformation are shown
in Figure 5.

The radial magnetic field vector ${\bf B}_r$ (which is pointing radially 
away from a magnetic charge $j$ located in the solar interior at 
$(x_j, y_j, z_j)$ is simply given by the difference of the Cartesian
coordinates from an arbitrary location $(x,y,z)$,
\begin{equation}
	{{\bf B}_{r} \over B_r} =  
	\left[ {x-x_j \over r_j}, {y-y_j \over r_j}, {z-z_j \over r_j}\right]
	= \left[ \cos_{r,x}, \cos_{r,y}, \cos_{r,z} \right] \ ,
\end{equation}
where $B_r$ is the absolute value of the radial magnetic field component 
$B_r(r_j, \theta_j)$ (Equation (34)), 
$r_j$ is the spatial length of the radial vector ${\bf r}_j$ (Equation (3)),
defining the directional cosines $\cos_{r,i}$ (for the 3D coordinates 
$i=x,y,z$) of the radial magnetic field vector ${\bf B}_r$.

The azimuthal component ${\bf B}_{\varphi}$ (with the absolute value
$B_\varphi(r_j, \theta_j)$ defined in Equation (35)) of the twisted 
magnetic field is orthogonal to the direction of the twist axis
${\bf R}$ (aligned with the local vertical), 
\begin{equation}
	{\bf R} = \left[ x_j, y_j, z_j \right] \ ,
\end{equation}
and the radial magnetic field component ${\bf B}_r$ (Figure 5),
and thus can be computed from the vector product of the two vectors
${\bf B}_r$ and ${\bf R}$, 
\begin{equation}
	  {{\bf B}_{\varphi} \over B_{\varphi}}
	= { {\bf R} \times {{\bf B}_r} \over |{\bf R} \times {\bf B}_r|} 
	= \left[ \cos_{\varphi,x}, \cos_{\varphi,y}, \cos_{\varphi_z} \right]\ ,
\end{equation}
which defines the directional cosines $\cos_{\varphi,i}$ of the azimuthal
component in the Cartesian coordinate system. The vector product
allows us also to extract the inclination angle 
$\theta_j$ between the radial magnetic field component ${\bf B}_r$ and the 
local vertical direction ${\bf R}$,
\begin{equation}
	\theta_j = \sin^{-1} \left( { |{\bf R} \times {\bf B}_r|
	\over |{\bf R}| \ |{\bf B}_r| } \right) \ .
\end{equation}

Finally, the total non-potential magnetic field vector 
${\bf B}=(B_x, B_y, B_z)$ is then the vector sum of the radial ${\bf B}_r$
and the azimuthal magnetic field component ${\bf B}_\varphi$, 
\begin{equation}
	\begin{array}{ll}
	B_x &= B_r(r_j, \theta_j) \cos_{r,x} 
		+ B_{\varphi}(r_j, \theta_j) \cos_{\varphi, x} \\
	B_y &= B_r(r_j, \theta_j) \cos_{r,y} 
		+ B_{\varphi}(r_j, \theta_j) \cos_{\varphi, y} \\
	B_z &= B_r(r_j, \theta_j) \cos_{r,z} 
		+ B_{\varphi}(r_j, \theta_j) \cos_{\varphi, z} 
	\end{array} \ ,
\end{equation}
with the directional cosines ($\cos_{r,i}$, $\cos_{\varphi,i}$, 
$\cos_{\theta,i}$) defined by Equations (38) and (40). 
This is a convenient parameterization that allows us directly to
calculate the magnetic field vector of the non-potential field 
${\bf B}_j=(B_x, B_y, B_z)$ associated with
a magnetic charge $j$ that is characterized with five parameters:
$(B_j, x_j, y_j, z_j, \alpha_j)$, where we define the force-free
$\alpha$-parameter from the twist parameter $b_j=2\pi N_{\rm twist}/l$
(Equation (7)) at the location of the twist axis ($\theta_j=0$), 
\begin{equation}
	\alpha_j= \alpha(\theta_j=0)  = {2 \ b_j} \ ,
\end{equation}
according to Equation (37).  

\subsection{ 	Superposition of Twisted Field Components 	}

The total non-potential magnetic field from all $j=1,...,N_{\rm m}$ magnetic
charges can be approximately obtained from the vector sum of 
all components $j$ (in an analog way as we applied in Equation (6) 
for the potential field), 
\begin{equation}
        {\bf B}({\bf x}) = \sum_{j=1}^{N_{\rm m}} {\bf B}_j({\bf x}) \ ,
\end{equation}
where the vector components ${\bf B}_j=(B_{x,j}, B_{y,j}, B_{z,j})$
of the non-potential field of a magnetic charge $j$ are defined in 
Equation (42), 
which can be parameterized with $5 N_{\rm m}$ free parameters 
$(B_j, x_j, y_j, z_j, \alpha_j)$ for a non-potential field, or with
$4 N_{\rm m}$ free parameters for a potential field (with $\alpha_j = 0$). 
Of course, the sum of force-free magnetic field vectors is 
generally not force-free, but we will prove in the following
(Equations (46) and (47)) that the sum of NLFFF solutions of the form of 
Equations (34)--(37)), which are force-free to second-order accuracy in 
$\alpha$ (or, more strictly, in $[b r \sin \theta]$), have the
property that their sum is also force-free to second-order in $\alpha$.
 
Let us first consider the condition of divergence-freeness.
Since the divergence operator is linear, the superposition of a number of
divergence-free fields is divergence-free also,
\begin{equation}
        \nabla \cdot {\bf B} = \nabla \cdot (\sum_j {\bf B}_j)
        = \sum_j (\nabla \cdot {\bf B}_j) = 0 \ .
\end{equation}
While the divergence-free condition is exactly fulfilled for a
potential field solution (Equation (4)), our quasi-forcefree approximation 
(Equations (34)--(37))
matches this requirement to second order in $\alpha$, as the insertion
of the solutions (Equations (34)--(37)) into the divergence expression 
(Equation (24)) shows. 
For a quantitative measure of this level of accuracy we can also
check numerical tests of the figure of merit (Section 3.3).

Now, let us consider the condition of force-freeness.
A force-free field has to satisfy Maxwell's equation (Equation (15)). 
Since we parameterized both the potential field and the non-potential
field with a linear sum of $N_{\rm m}$ magnetic charges, the requirement
would be, 
\begin{equation}
	\nabla \times {\bf B} =
	\nabla \times \sum_{j=1}^{N_{\rm m}} {\bf B}_j = 
	\sum_{j=1}^{N_{\rm m}} (\nabla_j \times {\bf B}_j) = 
	\sum_{j=1}^{N_{\rm m}} \alpha_j ({\bf r}) {\bf B}_j = 
	\alpha ({\bf r}) {\bf B} \ .
\end{equation}
Generally, these three equations of the vector ${\nabla \times \bf B}$
cannot be fulfilled with a scalar function $\alpha({\bf r})$ for a
sum of force-free field components, unless
the magnetic field volume consists of spatially separated 
force-free subvolumes. However, we can show the validity 
of the force-freeness equation (Equation (46)) to second-order accuracy in 
$\alpha$.
Note, that the nonlinear force-free parameter $\alpha$ is proportional
to $b$ (Equations (37) and (43)), which is defined in Equation (17), and thus 
we set 
second-order accuracy in $b$ equal to second-order accuracy in $\alpha$. 
The argument goes as follows. If we use spherical coordinates,
the NLFFF solution of the radial component is of zeroth order,
$B_r(r, \theta) \propto O(\alpha^0)$ (Equation (34)),
the azimuthal component is of first order, $B_\varphi(r, \theta) \propto
O(\alpha^1)$ (Equation (35)), 
and the neglected third component magnetic field component
is of second-order, $B_{\theta}(r, \theta) \propto O(\alpha^2)$ 
(as it can be shown by inserting $B_r$ and $B_\varphi$ into Equation (26).
The curl of the magnetic field (Equations (25)--(27)) is then of first order
for the radial component, $[\nabla \times {\bf B}]_r \propto \alpha B_r
\propto O(\alpha^1)$ (Equation (25)), to second order for the azimuthal 
component,
$[\nabla \times {\bf B}]_\varphi \propto (\alpha B_\varphi )
\propto O(\alpha^2)$ (Equation (27)), and the remaining third component 
is of third-order, $[\nabla \times {\bf B}]_\theta 
\propto (\alpha B_\theta ) \propto O(\alpha^3)$ 
(Equation (26)).  

Therefore, if we neglect second-order and higher-order terms, the
divergence-free condition (Equation (46)), which generally has three equations
for the three curl components, \eg $[\nabla \times B_r]_r$,  
$[\nabla \times B_\varphi]_\varphi$, $[\nabla \times B_\theta]_\theta$,
reduces to one single equation for the radial component,
$[\nabla \times B_r]_r$, which can be fulfilled with a scalar function
$\alpha({\bf r})$,
\begin{equation}
	\alpha({\bf r}) \approx {[\nabla \times {\bf B}]_r \over B_r} 
	= {[\nabla \times \sum_{j=1}^{N_{\rm m}} {\bf B}_j]_r  
	\over \sum_{j=1}^{N_{\rm m}} {\bf B}_{j,r} } 
	= {[\sum_{j=1}^{N_{\rm m}} \nabla \times {\bf B}_j]_r  
	\over \sum_{j=1}^{N_{\rm m}} {\bf B}_{j,r} } 
	= {\sum_{j=1}^{N_{\rm m}} \alpha_j B_j \over
	   \sum_{j=1}^{N_{\rm m}} B_j } \ .
\end{equation}
Thus, we expect that the force-freeness is fulfilled to second-order accuracy
$O(\alpha^2)$ (or strictly speaking $O(b^2)$). 
We will demonstrate the near force-freeness of simulated examples 
in the next section. 

\section{	Simulations and Tests 			}

We are now going to simulate examples of the analytical nonlinear 
force-free solutions in order to visualize the magnetic topology
and to quantify the accuracy of the divergence-free and force-free
conditions.

\begin{figure}
\centerline{\includegraphics[width=1.0\textwidth]{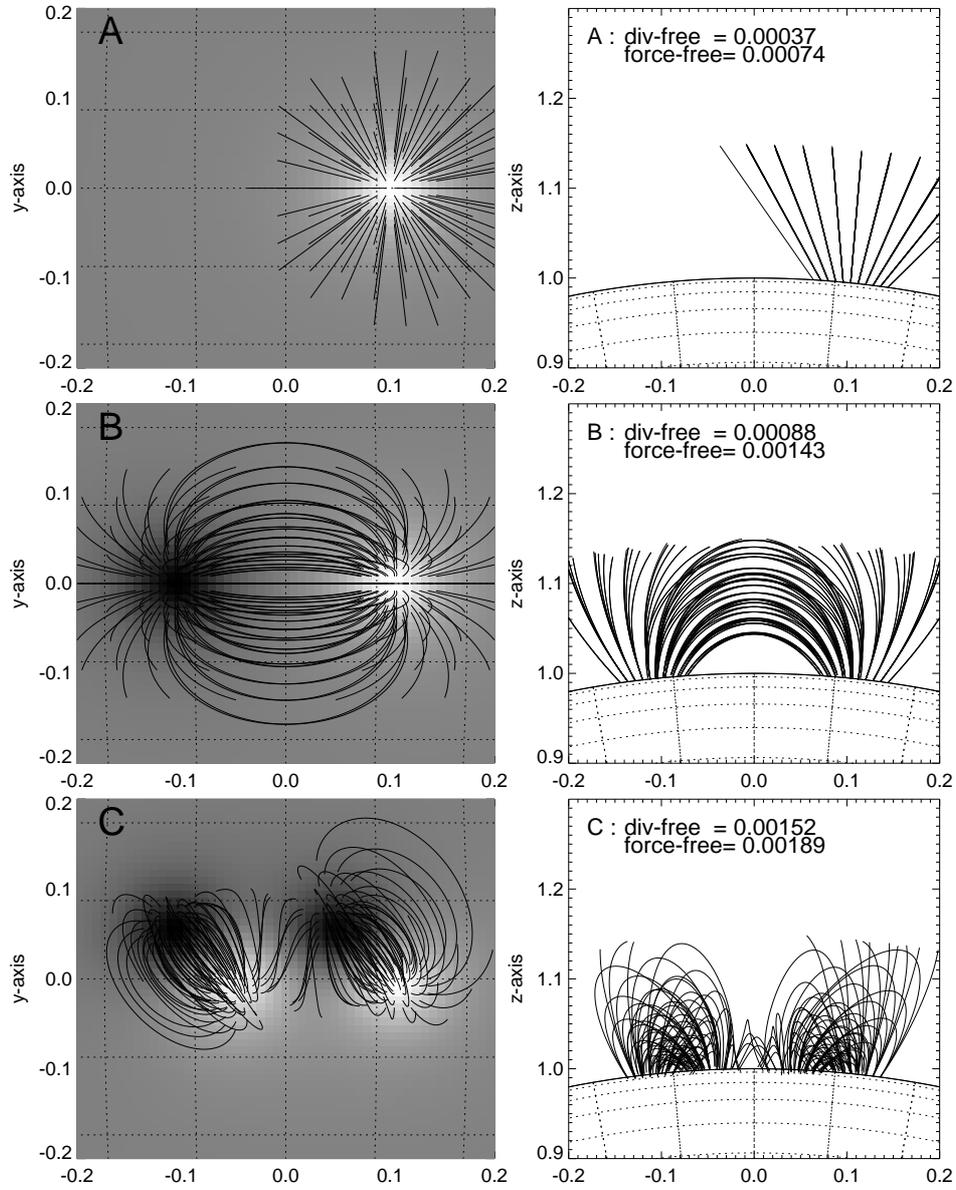}}
\caption{Simulations of three line-of-sight magnetograms (left) and
magnetic field lines projected into the $x-y$ plane (left) and into
the vertical $x-z$ plane (right). Thre three cases include:
(A) a single positive magnetic charge (first row), (B) a dipole
produced by two magnetic charges with opposite polarity (second row), 
and (C) a quadrupole configuration (third row). See parameters 
in Table 1. Only field lines with magnetic fields above a 50\%
threshold of the maximum field strength are shown.}
\end{figure} 

\begin{figure}
\centerline{\includegraphics[width=1.0\textwidth]{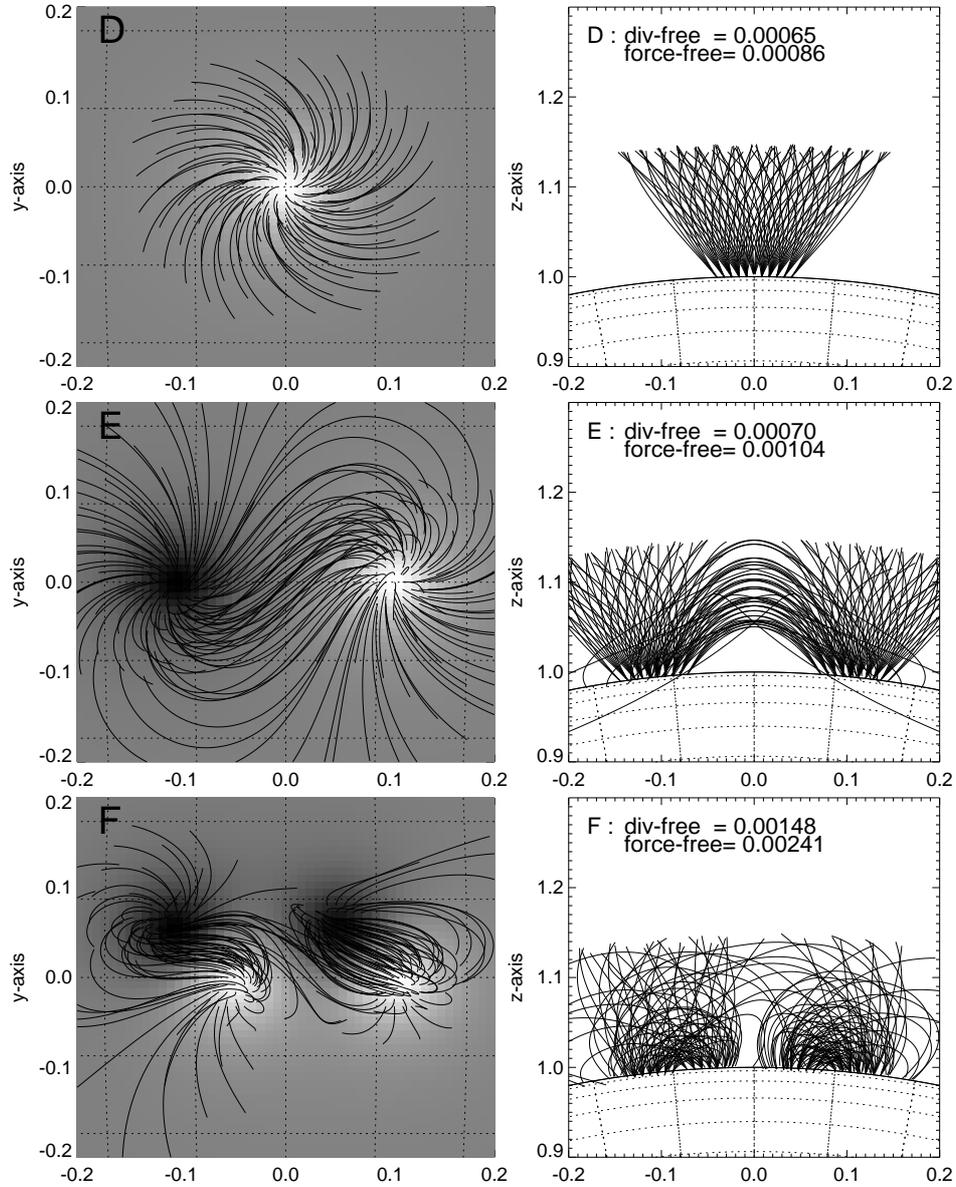}}
\caption{Simulations of three line-of-sight magnetograms (left) and
magnetic field lines of a non-potential model with currents are shown,
projected into the $x-y$ plane (left) and into
the vertical $x-z$ plane (right). The parameters of the three cases 
(D), (E), and (F) are identical to thoes of (A), (B), and (C), except
for the addition of electric currents.}
\end{figure}

\begin{figure}
\centerline{\includegraphics[width=1.0\textwidth]{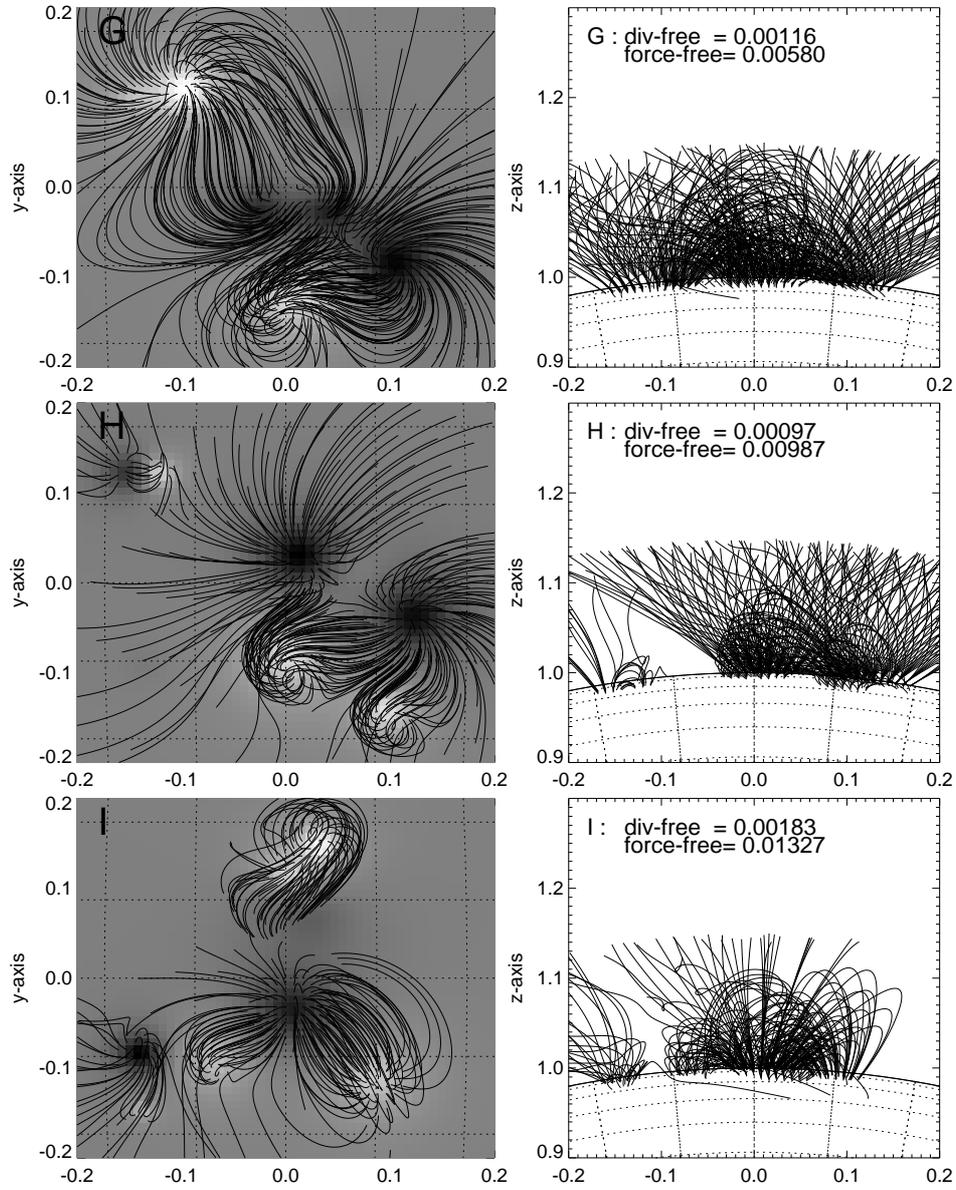}}
\caption{Simulations of three line-of-sight magnetograms (left) and
magnetic field lines of a non-potential model with currents are shown,
projected into the $x-y$ plane (left) and into
the vertical $x-z$ plane (right). The three cases (G), (H), and (I) 
have each $N_{\rm m}=10$ magnetic charges, with randomly chosen field strengths,
locations, and electric currents.}
\end{figure}

\subsection{	Numerical Examples			}

The simplest case is a single magnetic charge $j=1$, which we
illustrate as case A in Figure 6 (top row). We choose the following
parameters: a magnetic field strength of $B_1=1000$ G (gauss) at the solar
surface directly above the buried charge, the location 
$(x_1, y_1, z_1)=(0.1,0.0,0.95)$ for the buried charge, and
a number of zero twist $b_1=0$ for the potential field case.
We show the simulated line-of-sight magnetogram $B_z(x,y)$ in
Figure 1 (top left), which mimics an isolated sunspot. 
The pixel size of the magnetogram and the stepping size in the
extrapolation along a field line is $\Delta s = 0.004$ solar radii
(2800 km $\approx 4\arcsec$, corresponding to the pixel size of
SoHO/MDI magnetograms). We extrapolate the field lines for
every pixel that has a footpoint magnetic field strength above a 
threshold of 50\% ($>500$ G). The field lines point in radial
direction away from the center of the buried magnetic charge,
as it is expected for the potential field of an isolated sunspot
(and defined in Equation (1)). 

The next basic example is a magnetic dipole, which can be represented
in our model by a superposition of a pair of two magnetic charges
with opposite polarity, as sketched in Figure 1. The case B shown
in Figure 6 is simulated with equal, but oppositely signed magnetic 
field strengths ($B_1=1000$ G, $B_2=-1000$ G) at mirrored positions
($x_1=0.1, x_2=-0.1$), otherwise we used the same parameters as in case A 
($y_1=y_2=0.0, r_1=r_2=0.95, b_2=b_1=0.0$). The magnetic field lines 
mimic the familiar structure of a dipole, which is parameterized here
with 8 free parameters (in the potential case).

A quadrupolar configuration is simulated in case C (Figure 6, bottom),
with translational symmetry ($x_1=0.1, x_2=0.05, x_3=-0.05, x_4=-0.1;
y_1=0.1, y_2=0.05, y_3=0.1, y_4=0.05$), equal depths ($r_1=r_2=r_3=r_4=0.95$),
and alternating field strengths ($B_1=B_3=1000, B_2=B_4=-1000$ G).
The quadrupolar configuration shows essentially two bipoles, each one
with field lines that mostly connect within the same dipole domain, but
a few intermediate field lines actually connect from one to the other domain.

In Figure 7 we show the same three configurations as for the potential field
model (A, B, C of Figure 6), but add electric currents caused by twisting, 
corresponding
to $N_{\rm twist}=-0.5$ turns for the single charge (case D) or first dipole
(case E), and $N_{\rm twist}=1.0$ for the second dipole (case F),
defined for a loop length of $L=0.1 \pi$ solar radii.
These amounts of twist correspond to force-free $\alpha$-parameters
of $\alpha=2 \pi N_{\rm twist}/L=-10$ and $-20$ solar radius$^{-1}$
(\ie $\alpha=-1.43$ and $-2.86 \times 10^{-10}$ cm$^{-1}$). 
Comparing the potential (Figure 6) and non-potential cases (Figure 7)
shows clearly the differences that result from the presence of electric
currents. The force-free field lines of a sunspot become distorted into
spiral shapes (case D), the straight dipole becomes distorted into a
sigmoid shape (case E), and the quadrupolar configuration becomes also more
distorted with sigmoid-like structures (case F).

In Figure 8 we show a few more complicated cases (G, H, and I), consisting of 
$N_{\rm m}=10$ magnetic charges, with random values chosen in the magnetic
field range $-1000$ G $< B_j < +1000$ G, in positions 
$-0.15 < x_j < 0.15$ solar radii, $-0.15 < y_j < 0.15$ solar radii, 
$0.95 < r_j < 0.97$ solar radii, and random twist in the range
$-3 < N_{\rm twist} < +3$ per $L=0.1 \pi$ solar radii. 
The field lines displayed in Figure 8 demonstrate that a rich variety
of sigmoid-shaped dipoles and inter-connecting multi-pole configurations
can be generated with our quasi-force-free solutions, which mimic realistic
active regions observed in the solar corona. 

\begin{figure}
\centerline{\includegraphics[width=1.0\textwidth]{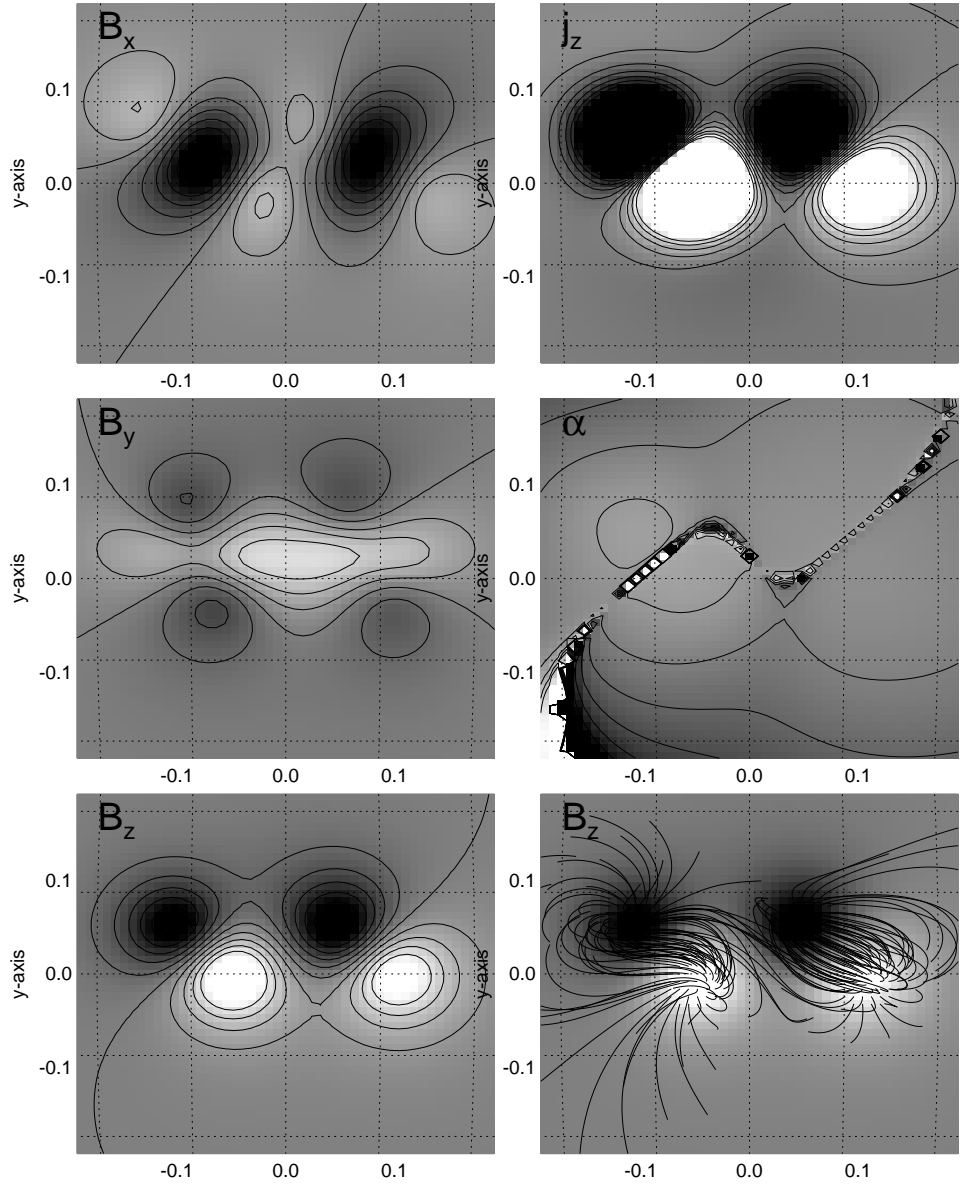}}
\caption{Maps of the magnetic field components $B_x(x,y), B_y(x,y), B_z(x,y)$
(left panels), the electric current density $j_z(x,y)$, and the force-free
$\alpha$-parameter (right panels).}
\end{figure}

\subsection{	Force-Free $\alpha$-Parameter and Electric Current Maps	}

In Figure 9 we show examples of various maps that can be generated to
visualize a 3D vector field solution, for the case F of a quadrupolar
configuration with currents. We show the following quantities
in the image plane $(x,y,z=1+\Delta s)$, which corresponds to an image
plane near the solar surface: The three magnetic field vector component
maps $B_x(x,y)$, $B_y(x,y)$, $B_z(x,y)$ (Figure 9, left panels,
the vertical electric current map $j_z(x,y)$ (Figure 9, top right panel),
the nonlinear $\alpha$-parameter $\alpha(x,y)$ (Figure 9, middle right panel),
and the LOS magnetogram $B_z(x,y)$ together with extrapolated field lines
(Figure 9, bottom right panel). The $B_z$ map shows most clearly the
locations of the four buried magnetic charges that form two dipolar or a
quadrupolar configuration. The magnetic polarization is also reflected 
in the $j_z$ and $\alpha$-map. The $B_z$ and the $\alpha$-map show also the
location of the neutral line, where numerical effects due to the limited
spatial resolution become visible. 

\begin{table}
\caption{Figures of merit for nine simulations of nonlinear force-free
field solutions, detailing the size of the 3D datacube, the number of magnetic
charges ($N_{\rm m}$), potential or non-potential model (P and NP), the
number of computed field lines $N_{\rm f}$, the divergence-freeness
$L_{\rm d}$, the force-freeness $L_{\rm f}$, and the computation times 
$t_{\rm CPU}$.}
\begin{tabular}{lrrrrrr}
\hline
Case    & Data cube             & Magnetic  & Field   & Divergence- & Force-	 & Computation\\
	&			& charges   & lines   & freeness    & freeness & time    \\
	&			& $N_{\rm m}$     & $N_{\rm f}$   & $L_{\rm d}$      & $L_{\bf f}$     & $t_{\rm CPU}$ (s)\\
\hline
A	& $51\times51\times37$  & 1 (P)     &  87     & 0.0004     & 0.0007     & 0.078 \\
B	& $51\times51\times37$  & 2 (P)     & 160     & 0.0009     & 0.0014     & 0.309 \\
C	& $51\times51\times37$  & 4 (P)     & 159     & 0.0015     & 0.0019     & 0.351 \\
D	& $51\times51\times37$  & 1 (NP)    &  87     & 0.0006     & 0.0009     & 0.083 \\
E	& $51\times51\times37$  & 2 (NP)    & 160     & 0.0007     & 0.0010     & 0.314 \\
F	& $51\times51\times37$  & 4 (NP)    & 159     & 0.0015     & 0.0024     & 0.414 \\
G	& $51\times51\times37$  & 10 (NP)   & 336     & 0.0012     & 0.0058     & 2.462 \\
H	& $51\times51\times37$  & 10 (NP)   & 302     & 0.0010     & 0.0099     & 1.764 \\
I	& $51\times51\times37$  & 10 (NP)   & 217     & 0.0018     & 0.0133     & 1.370 \\
\hline
\end{tabular}
\end{table}

\subsection{	Figures of Merit 			}

The degree of convergence towards a divergence-free
magnetic field model solution can be quantified by a measure that
compares the average divergence $\nabla \cdot {\bf B}$, which should be
close to zero, with the gradient $B / \Delta x$ of the magnetic field
over a reference length scale $\Delta x$, for instance a pixel of the computational
grid. The average deviation can then be defined by
(see also Wheatland \etal (2000) or Schrijver \etal (2006)),
\begin{equation}
	L_{\rm d} = {1 \over V} \int_V
	{|(\nabla \cdot {\bf B}) |^2
	\over |B / \Delta x|^2} dV \ .
\end{equation}
Similarly, the force-freeness can be quantified by the ratio of the
Lorentz force, $({\bf j} \times {\bf B}) = (\nabla \times {\bf B}) \times
{\bf B}$ to the normalization constant $B^2 / \Delta x$,
\begin{equation}
	L_{\rm f} = {1 \over V} \int_V
	{|(\nabla \times {\bf B}) \times {\bf B}|^2
	\over |B^2 / \Delta x|^2}  dV \ ,
\end{equation}
where $B = |{\bf B}|$. 

We calculated these figure of merit quantities for the nine cases simulated in 
Figures 6-9. The values are listed in each of the panels in Figures 6-8
and listed in Table I. The potential-field cases (A, B, and C) are found 
to have a figure of merit in the
range of $L_{\rm d}=0.0009 \pm 0.0006$ for the divergence-freeness, and
$L_{\rm f}=0.0014\pm0.0006$ for the force-freeness. 
The non-potential field cases (D, E, F, G, H, and I) 
have values in similar ranges of 
$L_{\rm d}=0.0009 \pm 0.0005$ for the divergence-freeness, and
$L_{\rm f}=0.0100\pm0.0080$ for the force-freeness. 
We find no tendency that this figure of merit depends on  
the number of magnetic charges or some
other model parameters. The fact that our quasi-force free analytical
solutions perform equally well as standard NLFFF codes described
in Schrijver {\it eq al.} (2006) tells us that the inaccuracy of the analytical
approximation is commensurable or even smaller than the numerical uncertainty 
of other NLFFF codes. However, since our analytical solution provides an
explicit formulation of nonlinear force-free fields, it can be
computed much faster than the standard NLFFF codes, and still provides
approximate solutions with acceptable accuracy (to second order).
The computation time of the analytical solutions
for the cases shown in Figures 6-8 amounts to about 
1 s (on a recent Mac computer:
Mac OS X, $2 \times 3.2$ GHz Quad-Core Intel Xeon, Memory 32 GB 800 MHz 
DDR2 FB-DIMM), while standard iterative NLFFF codes need 
several hours to converge to a single NLFFF solution. 

\section{	Discussion and Conclusions 		}

The coronal magnetic field has generally been computed by extrapolation
from lower boundary data in form of photospheric magnetograms 
$B_z(x,y,z=z_{ph})$ 
or vector-magnetograph data ${\bf B}(x,y)$, using a numerical extrapolation
algorithm that fulfills the conditions of force-freeness 
($\nabla \cdot {\bf B}$) and divergence-freeness $\nabla \times {\bf B}
= \alpha({\bf r}) {\bf B}$, where $\alpha({\bf r})$ is a scalar function
in space ${\bf r}$. These extrapolation algorithms are very computing-intensive,
because a good solution requires many iterations on a large computational 
3D-grid that has sufficient spatial resolution to resolve the relevant
magnetic field gradients. The accuracy of these numerical solutions depends
very much on the noise in boundary vector magnetic field data as well as
on deviations of photospheric fields from a force-free state. Recent
stereoscopic triangulation of coronal loops has demonstrated a considerable
mismatch between the extrapolated fields and the actual coronal loops,
which cannot easily be reconciled with extrapolation algorithms, since
they have only a very limited degree of freedom within the noise of the
boundary data. Moreover, since each NLFFF solution is very time-consuming 
to compute, these algorithms are not suitable for forward-fitting.

The forward-fitting of magnetic field solutions to observed data 
requires a faster algorithm to compute many NLFFF solutions for variable
boundary data or for coronal constraints as given by stereoscopic
3D reconstructions. The fastest computational way would be an explicit
analytical solution for the coronal field vectors ${\bf B}({\bf r})$ as
a function of some suitable parameterization of the boundary data or
coronal constraints. There exist some analytical solutions of nonlinear
force-free fields, such as a class of solutions in terms of Legendre
polynomials (Low and Lou, 1990), which is characterized by some spatial 
symmetry and has been used to test numerical
extrapolation algorithms (\eg DeRosa \etal 2009; Malanushenko \etal 2009).
However, to our knowledge, the class of analytical NLFFF solutions of 
Low and Lou (1990) has never been applied to forward-fitting of observed 
data, such as line-of-sight magnetograms, vector magnetograph 3D data, or 
to stereoscopically triangulated loops. Moreover, the special class of
NLFFF solutions derived in Low and Lou (1990) correspond to harmonics
of Legendre polynomials, which have a high degree of symmetry that does
not match realistic observations of active regions, and thus is not
suitable for forward-fitting to real data.

What we need to model observed solar magnetic data with high accuracy is:
(1) an explicit formulation of an analytical NLFFF solution;
(2) a parameterization of the NLFFF solution with a sufficient large
number of free parameters that can be forward-fitted to data and
converges close to observations; and (3) a fast computation algorithm
that can perform many interations without computing-intensive techniques.
Hence, such a project consists of developing a suitable analytical 
formulation first, and then to implement the analytical solutions into 
a forward-fitting code. In this paper we have
undertaken the first step. We started with a potential-field parameterization
in terms of $N_{\rm m}$ buried magnetic charges, which is defined by 
$4 N_{\rm m}$ free
parameters that can easily be extracted from an observed line-of-sight
magnetogram $B_z(x,y)$ with arbitrary accuracy, as demonstrated in two
recent studies (Aschwanden and Sandman, 2010; Aschwanden \etal 2012a).
The key concept of this potential-field representation is that an arbitrary
complex 3D magnetic field can be decomposed into a finite number of
elementary magnetic field components, where each one simply consists of
a quadratically decreasing radial field of a buried magnetic charge.
Divergence-freeness is conserved due to the linearity in the superposition
of elementary field components. In a next step we extended the elementary
potential-field component to a nonpotential-field component by adding
a uniform twist that can be parameterized by the force-free $\alpha$-parameter.
Such an elementary nonpotential field component requires five free parameters,
consisting of the four potential-field parameters plus the force-free 
$\alpha$-parameter. We derived an explicit analytical formulation of
the radial $B_r(r,\theta)$ and azimuthal field vector $B_\varphi
(r, \theta)$ that represents an approximative solution of the
divergence-free and force-free condition to second order ($\propto \alpha^2$).
This solution is very accurate for weakly non-potential fields and
converges to the potential field solution for $\alpha=0$. 
In analogy to the potential-field representation, we represent a
general non-potential field solution with a superposition of elementary
non-potential field components and prove that the divergence-freeness
and force-freeness is conserved to second-order accuracy in our
NLFFF approximation.

We calculated some examples of potential and non-potential fields
that mimic an isolated sunspot, a dipolar and a quadrupolar
configuration, as well as more complex multi-polar configurations.
The examples show that the magnetic field of arbitrary complex active 
regions can be represented with our parameterization. 
Increasing the force-free $\alpha$-parameter distorts circular field
lines into helical and sigmoid-shaped geometries. Our parameterization
allows one to compute either field lines (starting from arbitrary
locations), 3D datacubes of magnetic field vectors, of maps of
the force-free $\alpha$-parameter and electric current $j_z$ (Figure 9).
We tested the figures of merit for divergence-freeness and force-freeness,
which amount to $L_{\rm d} \lapprox 10^{-3}$ and $L_{\rm f} \lapprox 10^{-2}$. 
The examples demonstrate also the computing speed of this algorithm,
which amounts to the order of $\approx 1$ s for a computation grid that
encompasses a typical active region with the spatial resolution of MDI.
Thus, we envision that a full-fletched forward-fitting code can converge
within a few seconds to a few minutes, depending on the number of iterations
and number of magnetic field components.

Where do we go from here? The next step is the development of a
forward-fitting code that uses the magnetic field parameterization 
described here (see Paper II). 
We envision the applications to at least three different
sets of constraints, requiring three different versions of forward-fitting
codes: (i) line-of-sight magnetograms $B_z(x,y)$ and 3D coordinates 
$[x(s), y(s), z(s)]$ of stereoscopically triangulated loops;
(ii) line-of-sight magnetograms $B_z(x,y)$ and 2D coordinates 
$[(x(s),y(s)]$ of traced loops; and 
(ii) vector-magnetograph data $[B_x(x,y)$, $B_y(x,y)$, $B_z(x,y)]$.
The first application requires STEREO data, while the second one
can be obtained from any EUV imager (\eg AIA/SDO, TRACE, EIT/ SOHO).
The third application can be conducted with the new HMI/SDO data
and is equivalent to other NLFFF extrapolation codes without coronal
constraints, while the first two use coronal tracers 
and alleviate the force-free assumption of photospheric data.
We envision that these three applications will reveal insights into
a number of crucial questions in a novel way. 

There is a large number of physical problems and issues that can be addressed
with the anticipated forward-fitting code, such as:
(i) The force-freeness of the photosphere; 
(ii) the accuracy of NLFFF solutions;
(iii) the spatial distribution of electric currents in active regions;
(iv) the temporal evolution of currents before and during flares;
(v) the spatial distribution of current dissipation and coronal heating;
(vi) helicity injection;
(vii) the 3D geometry of coronal loops which is needed for hydrodynamic
modeling;
(viii) scaling laws of the volumetric heating function with other
physical parameters;
(ix) tests of the magnetic field strength inferred from coronal
seismology, \etc There exists hardly a phenomenon in the solar corona
that can be modeled without the knowledge of the coronal magnetic field.

\section*{Appendix A: The Gold-Hoyle Flux Rope}

A simple geometry of a force-free field structure is the Gold-Hoyle
flux rope (Gold and Hoyle, 1960), which consists of a curved
axis with helical field lines curved around the axis (Figure 10). 
While the stretched version of a flux rope with a straight twist axis
has the exact force-free solution of a uniformly twisted flux tube
(Section 2.2), the curved version of the Gold-Hoyle flux rope is
subject to curvature forces due to the gradient of the magnetic field
across the flux rope diameter and has a modified force-free solution.

\begin{figure}
\centerline{\includegraphics[width=0.5\textwidth]{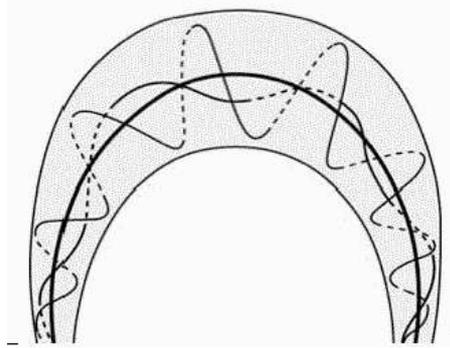}}
\caption{Cartoon of Gold-Hoyle flux rope.}
\end{figure}

\begin{figure}
\centerline{\includegraphics[width=0.8\textwidth]{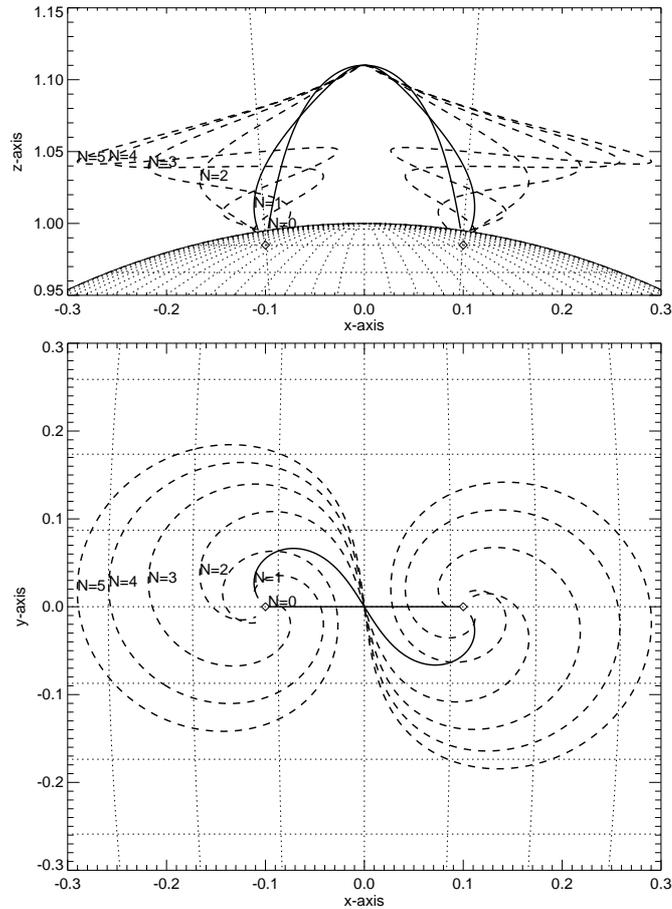}}
\caption{Dipolar field lines with various numbers of twisting turns:
$N=0$ (potential field line), stable sigmoid ($N=1$; solid line), 
and unstable sigmoids ($N=2,...,5$; dashed lines), according to our 
parameterization of point charges with twisted vertical axes.
Note that the limit of large twist numbers does not turn into
a Gold-Hoyle flux rope (Figure 10) with our parameterization.}
\end{figure}

In order to explore the limitations of our force-free field parameterization
we attempt here to model such a Gold-Hoyle flux rope. We
use the coordinates $(x_0,0,z_0)$ and $(-x_0,0,z_0$) with
$x_0=0.1$ and $z_0=0.985$ solar radii (marked with diamonds in Figure 11)
and extrapolate field lines $B(s)$ with our method, starting from the
apex position $(0,0,z_a)$ with $z_a=1.1$, for a set of six cases with
various force-free parameters $\alpha_1=\alpha_2$, where the $\alpha$'s
associated with the twist axis of each buried charge are defined by
$\alpha=2 \pi N_{\rm twist}/L$, with the loop length $L=2 \pi x_0=0.314$
and the number of twist turns $N_{\rm twist}=0, 1, ..., 5$ (indicated
with $N=0,...,5$ in Figure 11). The case $N=0$ corresponds to the
potential field case, yielding a coplanar elliptical loop shape.
The case $N=1$ represents a slightly twisted field line that has
a sigmoid shape and is a quasi-force-free solution. The cases
with $N=2,...,5$ are strongly twisted field lines and may be
less force-free, since the neglected $\alpha^2$ terms could be
significant.

Obviously we cannot reproduce the exact shape of the Gold-Hoyle
flux rope as shown in Figure 10 (with about seven twist turns) with our choice
of parameterization. The reason lies in the geometric constraints of the
twist axis, which is semi-circular in the case of the Gold-Hoyle model,
but consists of vertical twist axes in our parameterization. So, this
counter-example clearly demonstrates the limitations of our parameterization.
Nevertheless, although the cartoon with the Gold-Hoyle geometry is very popular,
especially for interplanetary flux ropes and CMEs, it is not clear whether
such Gold-Hoyle type geometries are found in loops in the lower corona,
and whether the Gold-Hoyle geometry corresponds to an exact force-free
solution. It is conceivable that the Sun exerts rotational stress
mostly in the photosphere (\ie rotating sunspots), which propagates
in vertical direction along the field lines, but does not necessarily
lead to a uniformly twisted circular flux tube as shown in Figure 10,
because the magnetic field drops rapidly with $r^{-2}$ with height
(for magnetic charges with small sub-photospheric depths), and thus the
magnetic stress is not uniformly distributed along a semi-circular
potential field line as envisioned in the Gold-Hoyle scenario.
However, for a case with a near-constant magnetic
field strength $B(s)$ along a potential field line, we would expect
a uniform twist as outlined in the Gold-Hoyle case.

On the other side, strongly twisted flux tubes with a twist larger
than about 1.25 full turns are unstable due to the kink instability
and may erupt, which is another reason why multiply twisted flux tubes
are unlikely to be found in active regions. Even Gold and Hoyle (1960)
found a critical twist number of $\Phi_{\rm twist} \lapprox 2.49 \pi$
$(N_{\rm twist} = \Phi_{\rm twist}/2 \pi \lapprox 1.25)$ above which no
equilibrium exists, which is also confirmed by recent MHD simulations
(\eg T\"or\"ok and Kliem, 2003). Thus, the Gold and Hoyle flux rope
case may not be relevant for modeling magnetic fields in stable
active regions. Nevertheless, more general parameterizations could be 
anticipated in future work, such as twist axes that follow potential 
field lines, rather than vertical axes, as used in our parameterization 
to minimize the number of free parameters.

\acknowledgements
We thank Anny Malanushenko for helpful discussions.
Part of the work was supported by
NASA contract NNG 04EA00C of the SDO/AIA instrument and
the NASA STEREO mission under NRL contract N00173-02-C-2035.

\section*{References} 

\def\ref#1{\par\noindent\hangindent1cm {#1}}

\small
\ref{Aschwanden, M.J.: 2004, {\sl Physics of the Solar Corona. An Introduction},
        Praxis Publishing Co., Chichester UK, and Springer, Berlin, 216.}
\ref{Aschwanden, M.J., W\"ulser, J.P., Nitta, N., Lemen, J.: 2008,
        \apj {\bf 679}, 827.}
\ref{Aschwanden, M.J.: 2009, \ssr {\bf 149}, 31.}
\ref{Aschwanden, M.J., Sandman, A.W.: 2010, \aj {\bf 140}, 723.}
\ref{Aschwanden, M.J., W\"ulser, J.P., Nitta, N.V., Lemen, J.R., DeRosa, M.,  
	Malanu\-shenko, A.: 2012a, \apj, submitted.}
\ref{Aschwanden, M.J., W\"ulser, J.P., Nitta, N.V., Lemen, J.R.: 2012b, \sp, 
	in press.}
\ref{Aschwanden, M.J., Malanushenko, A.: 2012, \sp, submitted, (Paper II).}
\ref{Boyd, T.J.M., Sanderson, J.J.: 2003, {\sl The Physics of Plasmas},
        Cambridge University Press, Cambridge, 102.}
\ref{Conlon, P.A., Gallagher, P.T.: 2010, \apj {\bf 715}, 59.}
\ref{DeRosa, M.L., Schrijver, C.J., Barnes, G., Leka, K.D., Lites, B.W.,
        Aschwanden, M.J., \etal: 2009, \apj {\bf 696}, 1780.}
\ref{Feng, L., Inhester, B., Solanki, S., Wiegelmann, T., Podlipnik, B.,
        Howard, R.A., W\"ulser, J.P.: 2007, \apjl {\bf 671}, L205.}
\ref{Gary, A., Alexander, D.: 1999, \sp {\bf 186}, 123.}
\ref{Gold, T., Hoyle, F.: 1960, \mnras 120, 89.}
\ref{Jackson, J.D.: 1962, {\sl Classical Electrodynamics},
	John Wiley and Sons, Inc., New York, 184.}
\ref{Klimchuk, J.A., Antiochos, S.K., Norton, D.: 2000,
	\apj 542, 504.}
\ref{Low, B.C., Lou, Y.Q.: 1990, \apj {\bf 352}, 343.}
\ref{Malanushenko, A., Longcope, D.W., McKenzie, D.E.: 2009,
        \apj {\bf 707}, 1044.}
\ref{Malanushenko, A., Yusuf, M.H., Longcope, D.W.: 2011,
        \apj {\bf 736}, 97.}
\ref{Metcalf, T.R., Jiao, L., Uitenbroek, H., McClymont, A.N., 
	Canfield,R.C.: 1995, \apj {\bf 439}, 474.}
\ref{Priest, E.R.: 1982, {\sl Solar Magnetohyrdodynamics},
        Geophysics and Astrophysics Monographs Volume 21,
        D. Reidel Publishing Company, Dordrecht, 125.}
\ref{Ruan, P., Wiegelmann, T., Inhester, B., Neukirch, T., Solanki, S.K.,
        Feng, L.: 2008, \aap {\bf 481}, 827.}
\ref{Sandman, A., Aschwanden, M.J., DeRosa, M., W\"ulser, J.P., 
	Alexander, D.: 2009, \sp {\bf 259}, 1.}
\ref{Sandman, A.W., Aschwanden, M.J.: 2011, \sp {\bf 270}, 503.} 
\ref{Schrijver, C.J., DeRosa, M.L., Metcalf, T.R., Liu, Y., McTiernan, J.,
	Regnier, S., Valori, G., Wheatland, M.S., Wiegelmann, T.:
 	2006, \sp {\bf 235}, 161.}
\ref{Sturrock, P.A.: 1994, {\it Plasma Physics. -- An Introduction to the
        Theory of Astrophysical, Geophysical and Laboratory Plasmas},
        Cambridge University Press, Cambridge, 216.}
\ref{T\"or\"ok, T., Kliem, B.: 2003, \aap {\bf 406}, 1043.} 
\ref{Wheatland, M.S., Sturrock, P.A., Roumeliotis, G., 2000, 
 	\apj {\bf 540}, 1150.}
\ref{Wiegelmann, T., Neukirch, T.: 2002, \sp {\bf 208}, 233.}
\ref{Wiegelmann, T., Inhester, B.: 2003, \sp {\bf 214}, 287.}
\ref{Wiegelmann, T., Lagg, A., Solanki, S.K., Inhester, B., Woch, J.:
        2005, \aap {\bf 433}, 701.}
\ref{Wiegelmann, T., Inhester, B.: 2006, \sp {\bf 236}, 25.}

\end{article}
\end{document}